\documentclass[aps,floatfix]{revtex4}
\usepackage{epsfig}

\newcommand{\beq}{\begin{eqnarray}}
\newcommand{\eeq}{\end{eqnarray}}
\newcommand{\bea}{\begin{eqnarray}}
\newcommand{\eea}{\end{eqnarray}}
\newcommand{\bean}{\begin{eqnarray*}}
\newcommand{\eean}{\end{eqnarray*}}
\newcommand{\ave}[1]{\left\langle #1\right\rangle}

\newcommand{\dd}{{\rm d}}
\newcommand{\e}{{\rm e}}

\begin{document}

\title{Internal
vs. External Conductivity of a Dense Plasma:
Many-particle theory and simulations}

\author{ H. Reinholz}
\affiliation{ University of Western Australia, School of Physics,
35 Stirling Highway, Crawley, WA 6009, Australia\\
phone +49 (0)381-498 2855, fax +49 (0)381-498 2857,
corresponding e-mail address: heidi@physics.uwa.edu.au}

\author{I. Morozov}
\affiliation{Institute for High Energy Densities of RAS,
IHED-IVTAN, Izhorskaya, 13/19, Moscow 127412, Russia}

\author{G. R\"opke and Th. Millat}
\affiliation{University of Rostock, FB Physik,
Universit\"atsplatz 3, D-18051 Rostock, Germany}
\date{\today}

\begin{abstract}

In the long-wavelength limit $k=0$, the response function has been
investigated with respect to the external and  internal fields which
is
expressed by the external and internal conductivity, respectively.
Molecular dynamics (MD) simulations are performed to obtain the
current-current correlation function and
the dynamical collision frequency which are compared with analytical
expressions.
Special attention is given  to the dynamical collision frequency and
the description of plasma oscillations in the case  of $k=0$.
The relation between the external and internal conductivity and to the current-current correlation function is analyzed.

\end{abstract}

\maketitle

PACS number(s):52.65.Yy, 52.25.Mq, 71.45.Gm, 52.27.Gr, 52.65.Vv

\section{Introduction}

The treatment of strongly correlated Coulomb systems is a challenge
for
many-particle theories.  It has applications in different fields such
as
dense ionic plasmas and the electron-hole plasma in excited
semiconductors.
Within a quantum statistical approach, the methods of equilibrium and
non-equilibrium Green functions have successfully been utilized  to
calculate the
properties of dense plasmas, see \cite{KKER}. However, a problem is
the
validity of perturbative approximations when using the Green
function approach for strongly correlated systems.

With increasing computer capacities, simulation techniques such as
molecular dynamics (MD) simulations have been developed to obtain
physical quantities from correlation functions, see
\cite{Hansen81a,Hansen76,Hansen87,DAN98,MNV01,Zwicknagel}. The MD
approach allows the application to large coupling
parameters. On the other hand, quantum effects are difficult to
include. This shortcoming is partially cured by considering
pseudopotentials which effectively take into account the
uncertainty principle by a short distance modification of the Coulomb
interaction within the range of the thermal wavelength, see
\cite{KKER}. More rigorous methods to include quantum effects are
wave packet MD simulations
\cite{zwicknagel} or path integral Monte Carlo calculations
\cite{millitzer}.

Other points are the finite particle number and the limited accuracy
when solving the equations of motion. The latter will not be
discussed any further.
The transition from a finite system to the thermodynamic limit of an
infinite system can be performed by periodic boundary conditions.
The total force on a given particle from all the other particles in a
basic cell, as well as from the infinite array of their periodic
images, can be obtained using the standard Ewald procedure
\cite{Hansen81a,Hansen76}.

In the present paper, the long-wavelength limit $\sigma(\omega) =
\lim_{k
\to 0}\sigma(k,\omega) $ of the dynamical conductivity is considered
for
a two-component plasma. According to the fluctuation-dissipation
theorem
(FDT), this transport quantity can be expressed in terms of
equilibrium
correlation functions, in particular the auto-correlation function
(ACF)
of the electrical current or the ACF of the electrical charge density.
In the literature \cite{Tkachenko}, see also \cite{Mahan}, the
internal as well as the external conductivity are introduced, relating
the electrical current density to the internal or the external
electrical field strength, respectively. We will present the
corresponding relations in the following Section II. An important
quantity related to the dynamical conductivity is the
dynamical collision frequency $\nu(\omega)$.
Analytical expressions can be derived in different approximations within a perturbative
approach, see
\cite{RRRW}.

Section III defines the current ACF in the context of MD simulations,
and the connection to the collision frequency is shown. While results
from MD simulations and analytical approaches for the structure
factor and other  frequency dependent quantities at finite wavenumber
$k$ are in good agreement, see e.g. \cite{Hansen81a,Zwicknagel}, we
will discuss the zero-wavenumber case of MD simulations which is
relevant  for the dielectric function $\epsilon(k=0,\omega)$ or the
dynamical conductivity $\sigma(\omega)$. Calculations are presented
without and including a mean field contribution which lead to the
external and internal conductivity, respectively.

Details of the MD simulations are reported in Section IV.
Results  for the current ACF  and the dynamical collision frequency
at  parameter values of a strongly coupled plasma are shown and
compared with results of the  analytical approach. The inclusion of a
mean field when performing MD simulations is considered in Section
V.   The controversy between the internal and external conductivity
in calculating the collision frequency is resolved.  Conclusions are
drawn in Section VI.

\section{Dynamical conductivity of the two-component plasma}

We consider a two-component fully ionized neutral plasma, such
as a H plasma consisting of electrons and protons, at temperature $T$
and density $n$ of each component.  The interaction is given
by the Coulomb potential, and the plasma is characterised by the
nonideality parameter $ \Gamma = e^2 (4\pi n_{e} / 3)^{1/3}
(4\pi\epsilon_0 k_BT)^{-1}$ and the degeneracy parameter $\Theta
\,=\,2m_ek_BT \hbar^{-2} (3 \pi^2 n)^{-2/3}$.  The linear response to
external perturbations in general is presented in various references,
see e.g. \cite{KKER,RRRW}. In the following, we will restrict
ourselves to relations which are relevant for further discussion.

Under the influence of
an external field $\vec E_{\rm ext}(\vec r,t) = \vec E_0^{\rm ext}
e^{i(\vec k \cdot \vec r - \omega t)}$ an electrical current density
$\langle \vec J_k\rangle^t$ is induced. The brackets $\langle \cdots \rangle^t$ indicate taking the statistical average with the thermodynamic equilibrium distribution.
If we consider the response in an isotropic systems, the $z$ axis can be
selected without loss of generality in such a way that $\vec E_0^{\rm
ext} = E_0^{\rm ext} \vec e_z,\,\,\vec k = k \vec e_z,\,\,\vec
J_k= J_k \vec e_z$.  The relationship between the
induced longitudinal current and the {\it external field} is given by
the response function $\chi(k,\omega)$.  Within  linear response
theory, $\chi(k,\omega)$ is related to the equilibrium correlation
function of the longitudinal electrical current density \cite{Hansen81a,RRRW}
\bea
  \chi(k,\omega) & = & -i\beta \Omega_0 {k^{2} \over \omega}
  \ave{J^{\rm long}_{k};J^{\rm long}_{k}}_{\omega+i\eta}
  \label{chi1}  \\
  & = &  -i\beta \Omega_0 {k^{2} \over \omega}
\int\limits_{0}^{\infty}\, \dd t
  \e^{i(\omega+i\eta)t} \ave{J^{\rm long}_{k}(t)J^{\rm long}_{k}},
  \label{chi2}
\eea
where $\Omega_0$ is the normalization volume and the limit $\eta \to 0$ has to be taken after the averaging over
the thermodynamic equilibrium. Since the longitudinal part of the
current density is related to the charge density according to the
balance equation (due to charge conservation), the longitudinal current ACF can
also be expressed in terms of the charge density ACF.

According to the FDT, the response function is related to the
dynamical structure factor or the longitudinal part of the dielectric tensor $\hat\epsilon(k, \omega)$ according to (c.f. \cite{Hansen81a,KKER,Mahan,Ichimaru})
\beq\label{epsdef}
\epsilon^{\rm long}(k, \omega) = 1 -  {\chi(k,
  \omega) \over \epsilon_0 k^2 + \chi(k, \omega)}
  = 1 - {1 \over \epsilon_0 k^2} \Pi^{\rm long}(k, \omega).
\eeq
The longitudinal polarization function $\Pi^{\rm long}(k, \omega)$ gives the
relation between the induced current and the {\it internal field} as
does the dynamical conductivity
\beq\label{Drude}
  \sigma^{\rm long}(k, \omega) = {i \omega \over k^2} \Pi^{\rm long}(k, \omega)
  = {\epsilon_0 \omega_{\rm pl}^2 \over -i \omega + \nu(k, \omega)}.
\eeq
It is also called the {\it internal conductivity} \cite{Tkachenko}.
In Eq.~(\ref{Drude}), the dynamical collision
frequency $\nu(k, \omega)$  is defined by a generalized Drude formula
where $\omega_{\rm pl} = (n e^2 / \epsilon_0 m_{ei})^{1/2}$ is the
plasma frequency and $m_{ei}$ the reduced mass.
The phenomenological Drude  model is found from the generalized Drude formula Eq. (\ref{Drude}) if the collision frequency is considered to be a real constant equal to the inverse of the relaxation time $\tau$ in momentum phase space.

In analogy to the internal conductivity, a so-called {\it external
conductivity} \cite{Tkachenko} can be introduced from the response
function (\ref{chi1})
\beq
\sigma_{\rm ext}(k, \omega) = {i \omega \over k^2} \chi(k, \omega)
 = \beta\Omega_0 \ave{J^{\rm long}_{k};J^{\rm long}_{k}}_{\omega+i\eta}.
  \label{kappa}
\eeq
This quantity is directly related to the longitudinal current ACF.  Note that it is
not the dynamical conductivity defined by Eq.~(\ref{Drude}). Instead, it is
related to the dynamical collision frequency in the following way
\beq
\sigma_{\rm ext}(k, \omega) = \frac{\epsilon_0\omega_{\rm pl}^2\omega}
  { -i(\omega^2-\omega_{\rm pl}^2)+\omega\nu(k, \omega) }\, .
  \label{KDrude}
\eeq

The transverse part of the dielectric tensor can also be related to a conductivity according to
\beq  \label{epsilon}
  \hat \epsilon (k, \omega) = 1 + {i \over \epsilon_0 \omega} \hat\sigma(k, \omega).
\eeq
The transverse conductivity is defined in analogy to the longitudinal (\ref{Drude}) as
\beq  \label{sigmatrans}
  \sigma^{\rm trans}(k, \omega) = {i \omega \over k^2} \Pi^{\rm trans}(k, \omega)
  = {\epsilon_0 \omega_{\rm pl}^2 \over -i \omega + \widetilde\nu(k, \omega)}.
\eeq
where $\widetilde\nu(k, \omega)$ is commonly called memory function \cite{Hansen81a}.
However, in this case, the Kubo-Greenwood formula \cite{Hansen81a,Hansen87,Ichimaru,Mahan, Kubo} relates the polarization function  directly to the transverse current ACF,
\beq \label{sigmatrans1}
  \sigma^{\rm trans}(k, \omega) =
   \beta\Omega_0 \ave{J^{\rm trans}_{k};J^{\rm trans}_{k}}_{\omega+i\eta}.
\eeq

Within a Green function approach, a diagram representation is possible
\cite{RRRW}.  In contrast to $\chi(k, \omega)$ and the transverse polarization function, which are given by
diagrams containing Coulomb interaction in any order,  the respective current ACF $\Pi^{\rm long}(k, \omega)$ is given only by the irreducible diagrams.   In the long-wavelength limit, transverse and longitudinal conductivies lead to the same response of the system,
\beq \label{kubo}
\lim_{k\rightarrow 0} \,\sigma^{\rm trans}(k,\omega)\,=\,\lim_{k\rightarrow 0} \,\sigma^{\rm long}(k,\omega).
\eeq


\section{Current Auto-Correlation Function}

Within MD simulations \cite{Hansen76,Hansen87,Mahan,Ichimaru}, the normalized current ACF
\beq
  K(t) = \frac{\ave{J_{k}(t)J_{k}}}{\ave{J_{k}^2}}
  \label{KMD}
\eeq
is calculated. Here, the long-wavelength limit ($k \to 0$) of the
current
\beq 
J_{k=0}(t)  = \frac{1}{\Omega_0}\sum_c\sum\limits_{i=1}^{N}
e_c
 v^z_{i,c}(t)
  \label{current}
\eeq
is considered, where $N$ is the number of electrons and singly
ionized
ions, and  $v^z_{i,c}$ is the speed in $z$-direction of the {\it i}th
particle of component $c$, denoted by $\{i,c\}$.  For convenience, we
will drop the index $k$ in the following.  Due to isotropy, the
normalizing factor is equal to
\beq
  \ave{J^2} =  \frac{e^2}{3\Omega^2_0} N \ave{v^2}
  = \frac{e^2}{\Omega^2_0} N\frac{k_{\rm B}T}{m}
  = \frac{\epsilon_0 \omega_{\rm pl}^2}{\Omega_0 \beta}.
  \label{avrcurrent}
\eeq
The Laplace transform of the current ACF reads
\beq
  \ave{J;J}_{\omega+i\eta}  = \ave{J^2}
  \int\limits_{0}^{\infty} \e^{i(\omega+i\eta)t} K(t)\,\dd t.
  \label{avrjk}
\eeq
On the basis of this quantity, two different results for the
conductivity
\beq
\sigma(\omega) =
  \epsilon_0\omega_{\rm pl}^2
  \int\limits_{0}^{\infty} \e^{i(\omega+i\eta)t} K(t) \dd t
  \label{sigmaMD}
\eeq
are derived depending on whether the current densities are considered to be long-wavelength limit of the  longitudinal or transverse case.

Firstly, within the transverse response, the Kubo-Greenwood formula (\ref{sigmatrans1})  is utilized. The conductivity (\ref{sigmaMD}) is then related to the memory
function  $\widetilde \nu(\omega)$
\cite{Hansen81a,Hansen87,Ichimaru,Kubo}  via the Drude like formula
(\ref{sigmatrans}) and we find
\beq \label{wrongDrude}
  {\widetilde \nu(\omega) \over \omega_{\rm pl}}
  = {\epsilon_0 \omega_{\rm pl} \over \sigma^{\rm trans}(\omega)}
  + i {\omega\over \omega_{\rm pl}}.
\eeq
If we assume a constant memory function (collision frequency)
$\widetilde \nu(\omega) = \widetilde \nu$, the Laplace transformation
of $\sigma^{\rm trans}(\omega)$ back to $K^{\rm trans}(t)$ using the functional dependence
given by Eq.~(\ref{sigmaMD}), leads to a monotonically decreasing
$K^{\rm trans}(t) =\exp( -\widetilde \nu t)$. This behaviour is  observed indeed
in simulations for $\Gamma \le 1$
\cite{Hansen81a,Hansen87,Ichimaru,Kubo,MNV01}.

Secondly, within  longitudinal response, we have to distinguish between the external
and the internal conductivity. Inserting Eqs. (\ref{avrjk}) and
(\ref{avrcurrent}) into (\ref{kappa}), this implies that expression (\ref{sigmaMD}) is the
external conductivity. The internal conductivity can be calculated via
\beq\label{sigkap}
\sigma^{\rm long}(\omega) = {\sigma_{\rm ext}(\omega) \over 1 -
  i \sigma_{\rm ext}(\omega)/(\epsilon_0\omega)}.
\eeq
 and due to the generalized Drude formula (\ref{Drude}) the collision
frequency is, in contrast to (\ref{wrongDrude}),
  \beq \label{nukap}
{\nu(\omega)\over \omega_{\rm pl}}= {\epsilon_0\omega_{\rm pl} \over
  \sigma_{\rm ext}(\omega)} + i \left( {\omega\over \omega_{\rm pl}} -
  {\omega_{\rm pl} \over \omega} \right).
\eeq

Using a constant collision frequency $\nu(\omega) = \nu$ in the
respective relationship (\ref{KDrude}) for the external conductivity,
we find for the longitudinal current ACF via a Laplace transformation
\beq
  K^{\rm long}(t) = \exp\left\{ -\frac{\nu}{2}\,t \right\}
  \left[ -{\nu \over 2z}\sin(zt)+\cos(zt) \right],\quad
  z = \sqrt{\omega_{\rm pl}^2-{\nu^2\over 4}}.
\eeq
This shows that an oscillating behaviour is expected for the ACF. The
oscillation frequency tends to $\omega_{\rm pl}$ in the limit $\nu \to 0$.

If in the long-wavelength limit both $\nu(\omega)$ and $\widetilde\nu(\omega)$ coincide, the current ACF for the longitudinal and transverse response cannot be identical.
In the following Sections IV,V we will resolve the contradiction
between the internal conductivity as obtained from the current ACF
according to (\ref{sigkap}) and (\ref{kappa}) and the transverse conductivity obtained from the current ACF according to (\ref{wrongDrude}) and (\ref{sigmatrans1}).


\section{Simulation technique}

In the  MD simulation scheme, the Newtonian equations of
motion are solved for a system consisting of $N$ singly charged ions
and $N$ electrons exerting Coulomb forces on each other.  The
{\it i}th particle of component $c$ shall be denoted as $\{i,c\}$.
This is a classical treatment where the trajectories of each particle
are determined.  The original Coulomb interaction can be replaced by
a
pseudopotential, where the short-range part of the interaction is
modified reflecting the quantum character of the interaction.  A
systematic derivation of a pseudopotential which reproduces the
equilibrium properties has been given by Kelbg,
see~\cite{KKER,CKelbg}
on the basis of the Slater sum.  In particular, we use the so-called
``corrected Kelbg'' potential~\cite{CKelbg}:
\beq\label{CorKelbg}
  V_{cd}(r) = \frac{e_c e_d}{4 \pi \epsilon_0 r} \left[
    F\left({r \over \lambda_{cd}}\right) - r \frac{k^{}_B T}{e_c e_d}
    \tilde{A}_{cd}(\xi_{cd})
    \,\exp\left(- \left({r \over \lambda_{cd}}\right)^2 \right)
  \right],
\eeq
where
\bean
& & \hspace*{-20pt}
    \lambda_{cd}={\hbar \over \sqrt{2m_{cd} k^{}_B T}}  \quad
    {1 \over m_{cd}} ={1 \over  m_c} + {1 \over m_d},\quad \xi_{cd}
    = -{e_c e_d \over k^{}_B T \lambda_{cd}}, \\
& & \hspace*{-20pt}
    F(x) = 1 - \exp(-x^2) + \sqrt{\pi} x (1 - {\rm erf}(x)), \\
& & \hspace*{-20pt}
    \tilde{A}_{ee}(\xi_{ee}) = \sqrt{\pi} |\xi_{ee}| + \ln\left[
    2\sqrt{\pi} |\xi_{ee}| \int\limits_0^{\infty}
     \frac{y \exp(-y^2)\, dy}{\exp(\pi|\xi_{ee}|/y) - 1}
 \right], \\
& & \hspace*{-20pt}
   \tilde{A}_{ei}(\xi_{ei}) = -\sqrt{\pi} \xi_{ei} + \ln\left[
   \sqrt{\pi} \xi_{ie}^3 \left( \zeta(3) + \frac14 \zeta(5)\xi_{ie}^2
   \right)
   \vphantom{\int\limits_0^{\infty}} \right.
     \left. +\; 4\sqrt{\pi} \xi_{ei} \int\limits_0^{\infty}
     \frac{y \exp(-y^2)\, dy}{1 - \exp(-\pi\xi_{ei}/y)}
 \right].
\eean
where $\xi(n)$ are the Riemann-Zeta functions.  This interaction
potential corresponds to the Coulomb potential at large distances and
provides the exact value of the Slater sum and its first derivative
at
$r=0$.

Initially, all the particles are gathered in a cubic box with the
edge
size $L$.  The number of particles $N$ in this basic cell is obtained
from a given mean plasma density $n$ via $N = n L^3$.  To simulate an
infinite homogeneous plasma, images of this charge-neutral basic cell
are considered shifting the basic cell by integer multiples of $L$
in
different directions.  This extended system has a constant mean
plasma
density $n$.  Artefacts may occur due to the periodicity of the
particle positions, but they are suppressed if the basic cell size is
increased.

The dynamics of both electrons with charge $-e$, mass $m_{e}$ and
ions
with charge $e$, mass $m_{i}$ is considered.  Because of the
continuous expansion of such plasma, the nearest image method is
applied to the force calculation procedure.  Here, the force $ \vec
F_{i,c}= \vec F_{i,c}^{\,\,\rm short} +
\vec F_{i,c}^{\,\,\rm long}$ on a particle $\{i,c\}$  is considered
to consist
of two contributions.  The interaction forces between particle
$\{i,c\}$ and the nearest neighbour images of all other particles
found in the basic cell centered around the position $\vec r_{i,c}$
of
the considered particle $\{i,c\}$ is the short-range contribution $
\vec F_{i,c}^{\rm short}$.  The contribution $
\vec F_{i,c}^{\rm long}$ is originated from the remaining images,
which
are not in the basic cell.

The short-range part of the force is calculated as
\beq\label{IntForce}
  \vec F_{i,c}^{\,\,\rm short} =
   \sum_d \sum\limits_{j (\neq i)}^{N}
   \vec F_{cd}(\vec r^{\,\,\rm n.n.}_{j,d} -  \vec r^{}_{i,c}), \quad
  \vec F_{cd}(\vec r) = - \frac{\vec r}{r} {d V_{cd}(r) \over dr} .
\eeq
The time argument $t$ is suppressed.  According to this method it is
assumed that the particle $\{i,c\}$ doesn't interact with  original
particles which at large $t$ may be found far away due to the
motion in space beyond the basic cell, but with their next
neighbours' images obtained by
periodically shifting their coordinates into the basic cell centered
around the particle $\{i,c\}$.  Thus, the position of each original
particle $\vec r_{j,d}$ is replaced by the position of an image
${\vec r_{j,d}}'$
\beq
  r^{\,\,\rm n.n.,\alpha}_{j,d} = r^\alpha_{j,d} - mL, \quad
  \left| r^{\,\,\rm n.n.,\alpha}_{j,d} - r^\alpha_{i,c} \right|
  \le \frac{L}2,
\eeq
where $\alpha = x,y,z$ and $m$ is an integer. It should be noted that
this procedure is repeated for each particle at $\vec r_{i,c}$.
This method implies that each particle is always surrounded by $2N-1$
other particles with a constant mean density and the plasma is
homogeneous in scales larger than the simulation cell.

The forces $ \vec F_{i,c}^{\,\,\rm long} $ due to the interaction with
images outside the basic cell centered around the position $\vec
r_{i,c}$ of the particle $\{i,c\}$ are treated in a different way.
If
the dimension $L$ of the basic cell is large in comparison to the
screening length, the contributions of all images except the nearest
one can be neglected.  In particular, this is justified in the case
of a
nonideal plasma where the effective interaction potential decreases
exponentially with distance due to screening.  The influence of the far images can be taken into account considering Ewald sums.  They are expected to
give only a small contribution to $\vec
F_{i,c}^{\,\,\rm short}$ provided $N$ is high enough.
They are not relevant with respect to our considerations.

For explicit MD simulations,  we consider a model plasma consisting
of  singly charged ions and electrons with density $n = 3.8 \times
10^{21}$ cm$^{-3}$ at a temperature of $T=33\:000$ K. This
corresponds to recent experiments in dense xenon plasmas
\cite{Refl03}.  The plasma parameters introduced in Sec.~II take the
value $\Gamma = 1.28, \,\, \Theta = 3.2$.  It is a nondegenerate,
strongly coupled plasma.  The computations of the current ACF for the
ion-electron mass ratios $m_i/m_e = 1836$ and $m_i/m_e = 100$ show no
considerable difference.  Thus the ratio $m_i/m_e = 100$ is selected
for better convergence when averaging over the configurations of
ions.
The total number of particles $N = 250$ was found to be enough for
$\Gamma \approx 1$.  Further increase of the number of particles ($N
=
400$) does not affect any simulation results including the mean
interaction energy, equilibrium correlation functions and others.
The
equilibrium state of the plasma at the given temperature was obtained
using a special procedure described in~\cite{DAN98}.

The current ACF is calculated directly from the velocities of the
particles in subsequent moments of time according to Eqs.
(\ref{KMD})
and (\ref{current}), where $\Omega_0=L^3$ with $L$ the length of the
basic cell.  The averaging of the ACF is performed over
$(1-5)\cdot 10^5$ initial configurations.  These configurations are
obtained from a long MD trajectory at different time moments.  As
shown in~\cite{MNV01}, two configurations are statistically
independent if they are taken at times separated by the dynamical
memory time.  In our case about $5\cdot 10^3$ initial configurations
are already fully statistically independent for electrons.  The
dynamical
memory time for ions increases with the ion mass~\cite{MNV01}.  Thus
the smaller mass ratio the better averaging for ions is obtained.

\begin{figure}[ht]
\begin{center}
  \includegraphics[width=0.5\linewidth]{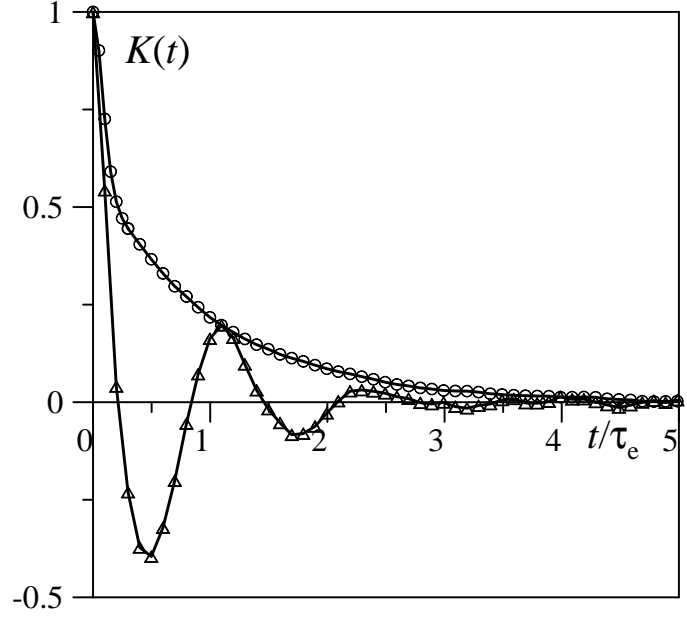}
  \vspace*{-15pt}
\end{center}
\caption{Current auto-correlation function (ACF) for $\Gamma=1.28$,
$m_i/m_e=100$; total number of averages $5\times 10^{5}$; MD
trajectory length of $2.5\times 10^{4}\tau_e$,
$\tau_e=2\pi/\omega_{pl}$
--  period of electron plasma oscillations: MD simulations without
(circles) and including (triangles) an additional mean-field term in
the equations of motion.
  \label{figjj}}
\end{figure}

Results are shown in Fig.~\ref{figjj} with circles. The relatively small ion-electron mass
ratio ($m_i/m_e=100$) was chosen for computational reasons since the
calculation with greater mass ratio shows exactly the same results for the
current ACF. The current ACF $K(t)$ decreases monotonously as it was also
obtained in previous MD simulations \cite{Hansen81a,Hansen87,MNV01}. It
indicates that the conductivity obtained numerically from $K(t)$ according to (\ref{sigmaMD}) should be treated as the transverse conductivity (\ref{sigmatrans}).
The dimensionless dynamical conductivity $\sigma(\omega)/ (\epsilon_0
\omega_{\rm
pl})$ is shown in Fig.~\ref{figKw} with circles.   As $\omega \rightarrow 0$, the real part  has a finite value  and the imaginary part vanishes, as expected from
Eq.~(\ref{wrongDrude}). According to the latter expression, we then
deduct a memory function or  collision frequency
$\widetilde\nu(\omega)$ as shown in Fig.~\ref{figNu} with circles.

Details of different approximations for the dynamical collision
frequency within a generalized linear  response theory can be found
in \cite{RRRW}.  The dynamical collision  frequency in Born
approximation with respect to the statically
screened potential (Debye potential) taken in the non-degenerate case
and within the long-wavelength limit, is given here
\beq \label{born}
  \nu^{\rm Born}(k=0,\omega) =-i g\,n\,  \int_0^\infty dy
  {y^4 \over 1 +{\bar n^2 \over y^4}}
  \left[\widetilde V(q)
    {16 m_e k_BT \Omega_0 \epsilon_0 \over e^2 \hbar^2} \right]^2
  \int_{-\infty}^\infty dx e^{-(x-y)^2} {1
    - e^{-4 xy} \over xy (xy-\bar \omega -i \eta)}\,\,,
\eeq
where
\beq
  q={y \over \hbar}\sqrt{16 m_e k_BT}\, , \quad
  \bar n = {\hbar^2 n e^2 \over 8 \epsilon_0 m_e (k_BT)^2}\, , \quad
  g = {e^4 \beta^{3/2} \over 24 \sqrt{2} \pi^{5/2} \epsilon_0^2
    m_e^{1/2}}\, , \quad
  \bar \omega = {\hbar \omega \over 4 k_BT}.
\eea
In the case of the Fourier transform of the Coulomb interaction
$\widetilde V(q) = e^2/(\Omega_0 \epsilon_0 q^2)$ the square brackets
become $1/y^2$.

\begin{figure}[ht]
\begin{center}
  \includegraphics[width=0.5\linewidth]{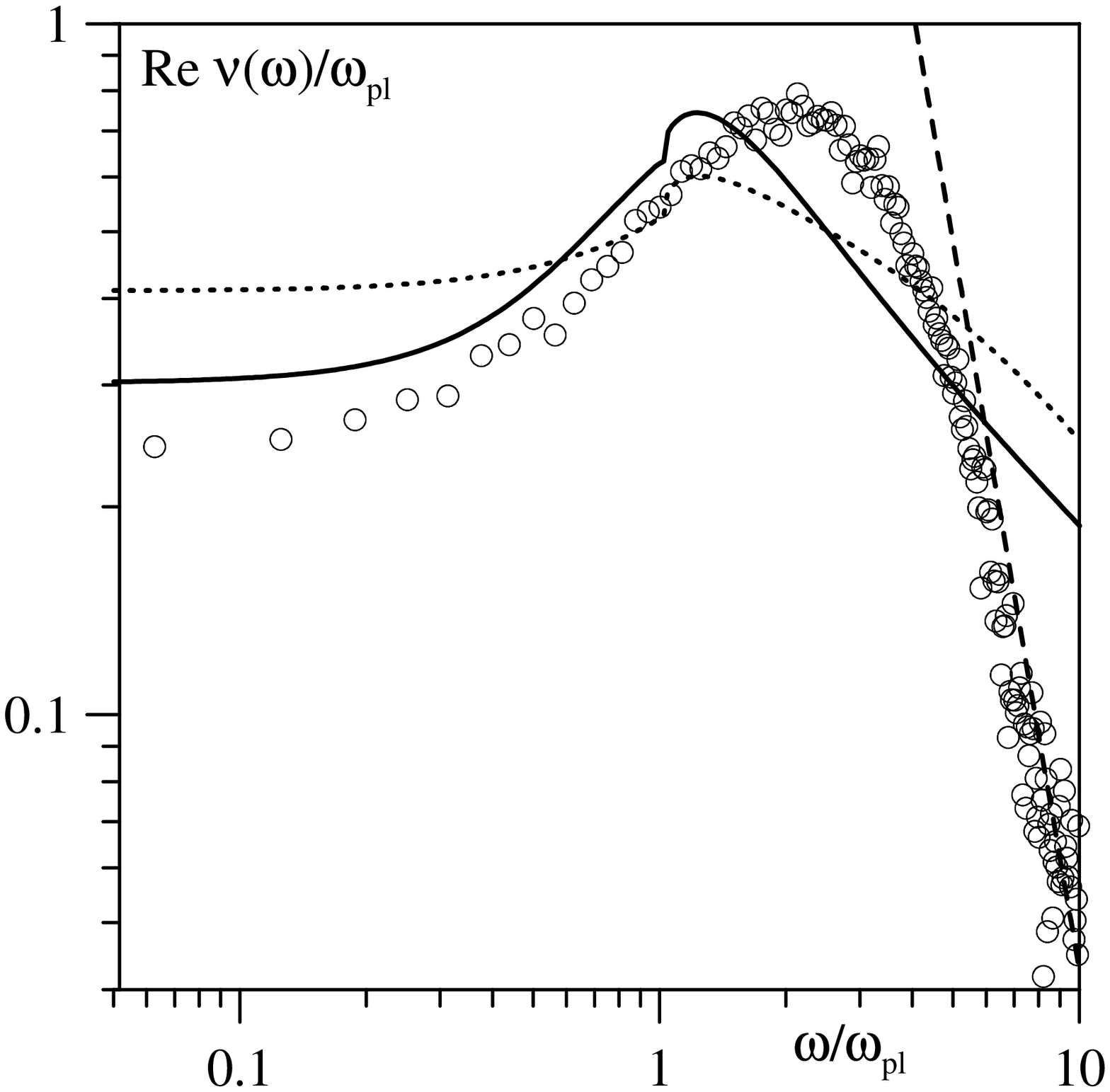}
  \includegraphics[width=0.5\linewidth]{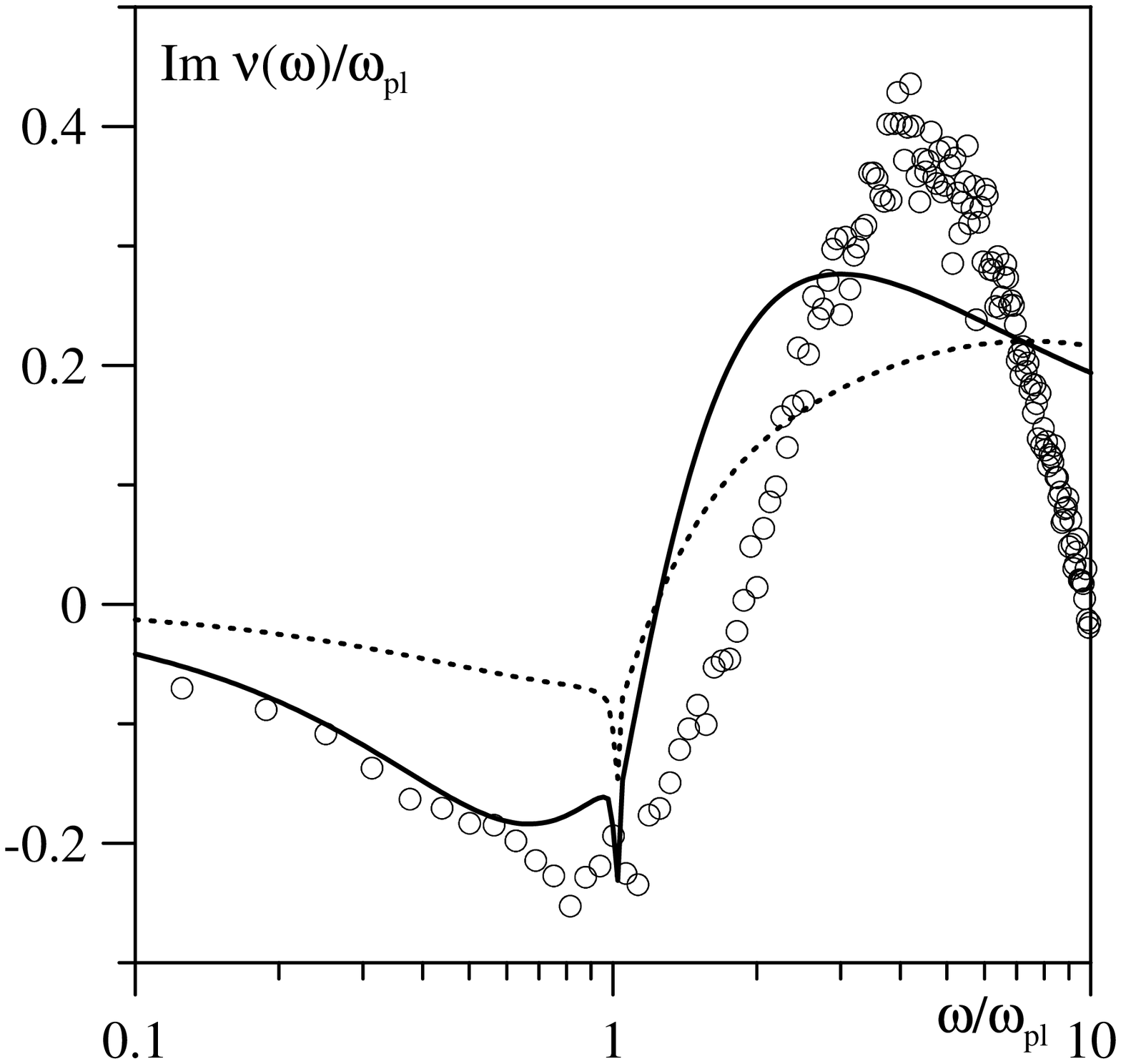}
  \vspace*{-15pt}
\end{center}
\caption{Dynamical collision frequency within different methods;
points -- MD simulations; analytical approximations:
 dotted line  -- Born approximation
Eq.~(\protect\ref{born}) with Coulomb potential,
 solid line --
same approach including dynamically screening and strong collisions
(T matrix)
and higher moments via renormalization factor \cite{RRRW},
dashed line -- high frequency asymptote for Born approximation
Eq.~(\protect\ref{born}) with corrected Kelbg potential.
  \label{figNu_lg}}
\end{figure}

We will now compare the MD simulations with this analytical treatment
of the dynamical collision frequency within perturbation theory, see
Figs.~\ref{figNu_lg}.  Firstly, we consider a system with statically
screened Coulomb interaction $\tilde V(q) = e^2/(\Omega_0
\epsilon_0 q^2)$ according to Eq.~(\ref{born}). The results are
presented  as dotted line.  The Born approximation can be improved by taking
into
account the effects of dynamically screening, strong collisions (T
matrix) and higher moments by introducing a renormalization factor
\cite{RRRW} in the generalized Drude formula Eq.~(\ref{Drude}).
This
approximation is shown as solid line.  Details of the calculation are
given in \cite{RRRW}.  It can be seen that both real and imaginary
part are in good agreement with the simulation results for
$\omega<\omega_{\rm pl}$.  This means that in this region the quantum
mechanical treatment of the Coulomb potential and the classical
simulations based on the corrected Kelbg potential are consistent.

At frequencies $\omega\gg\omega_{\rm pl}$ the asymptotic expansion of
the
analytical expression for the collision frequency is possible using
the
Fourier transform of the corrected Kelbg potential (\ref{CorKelbg})
\beq\label{FCorKelbg}
  \widetilde V_{cd}(q) = \frac{e_c e_d \lambda_{cd}}{ \epsilon_0
\Omega_0
    q} \left[ {\sqrt{\pi} \over \lambda^2_{cd} q^2}  {\rm Erfi}({
      \lambda_{cd} \over 2} q)e^{-{ \lambda^2_{cd} \over 4} q^2} - {
      \lambda^2_{cd} k_BT \pi^{3/2} \epsilon_0 \over e_ce_d }
    \tilde{A}_{cd}(\xi_{cd})\,q\,e^{-{ \lambda^2_{cd} \over 4} q^2}
\right].
\eeq
For the high frequency behaviour of the real part is found ${\rm
Re}\,\nu(\omega)\sim \omega^{-3.5}$ which is given in
Fig.~\ref{figNu_lg} as dashed line.  There is good agreement between  the simulation
data and the analytically derived high frequency behaviour.  The
presented
analytical treatment was also confirmed by MD calculations of the
dynamical structure factor at finite $k$ in~\cite{Zwicknagel} where
the Deutsch potential was used.

\section{Longitudinal conductivity}

We now investigate the evaluation of the longitudinal conductivity by MD simulations. The current ACF $K^{\rm trans}$ discussed in the previous Section cannot be taken since this current ACF yields the correct collision frequency only if the external conductivity is related to a Drude ansatz. However, this is not consistent. Instead  we have to derive the internal conductivity, from which a collision frequency can be obtained via the Drude formula. Therefore, the current ACF $K^{\rm long}$ has to be calculated differently than the ACF $K^{\rm trans}$. It will be shown how to obtain the longitudinal current ACF in the long-wavelength limit. However,
we note that for finite wavevector $k$ excellent agreement  for the dynamical structure factor from MD simulation and analytical expressions has been found \cite{Zwicknagel}. The condition $k>2\pi/L$ means that any charge density wave occurs already within the basic simulation cell and the corresponding mean electric field is accurately taken into account. The limit $k\rightarrow 0$ is not trivial. For any small $k$, the system is nearly homogeneous, but charge densities (or surface densities)  are present at large distances, which can also be considered as a mean field.

For this, we follow the procedure to construct an infinite system by
periodic images of a basic cell. We consider this as a limiting case
of a finite number of images. Denoting the images in $z$-direction by
$N_{\rm images}$, then a surface of our system is obtained at
$z_-=-(2N_{\rm images}+1) \cdot L/2$ and $z_+=(2N_{\rm images}+1)\cdot L/2$.
When considering the force calculation procedure, there are
contributions to the forces originating from a surface charge
density. This occurs if positive and negative charges are moving at
different rates across the surface of the basic cell.  The
introduction of a finite number of images compensates this effect at
the interfaces, but not at the surface of the whole system including
all the images. A large dipole moment follows  connected with a
finite polarization of the system. This surface charge density will
produce an electrical field which has to be taken into account even in the limit
when the number of images goes to infinity.  If the surface is far away, it
produces a homogeneous electrical field $\vec E(t)$ within the
simulation box.  Following this reasoning, it is necessary to include a mean field in the long-wavelength limit as shown below. As a consequence, plasma oscillations are obtained in the current ACF.

On the macroscopic level, the Maxwell equations relate this mean field $\vec E(t)$ to the average current density $\vec J(t)$, which is  oriented in $z$-direction according to the conventions in Sec.~II,
\beq\label{E-j}
  {d \vec E(t) \over dt} = - {1 \over \epsilon_0}\langle \vec J(t)\rangle.
\eeq
Taking the current density according to Eq.~(\ref{current}) as an average over the basic simulation cell and the initial condition $ \vec E(0) = 0 $, the integration of  Eq.~(\ref{E-j}) leads to
\bea\label{E-P}
& & \vec E = 
 \frac{1}{L^3}\left( -e \sum\limits_{i=1}^{N} \vec r_{i,e}
  + e \sum\limits_{i=1}^{N} \vec r_{i,i} \right)
  \label{TotPolar}
\eea
In this approach, the long-range interaction forces are given by
 $ \vec F_{i,c}^{\,\,\rm long}(t) = e_c \vec E(t)$.
In particular, the equation of motion for an electron includes two
parts
\beq\label{EqMotion}
  m_{e} {d \vec v_{i,e} \over dt} = \vec F_{i,e}^{\,\,\rm short} -
e\vec E.
\eeq
The interaction forces $\vec F_{i,e}^{\,\,\rm short}$ originate
from close partners in the Debye sphere within the basic cell.  It is
fluctuating around a nearly zero mean value.  Nevertheless, the
amplitude of these fluctuations are much higher then the fluctuations
of $e\vec E$.

In the  MD method, if no mean field term is taken into
account, the total energy
\beq \label{TotEnergy}
  {\cal E}_{\rm tot} = {\cal E}_{\rm pot} + {\cal E}_{\rm kin}
  = \frac 12 \sum\limits_{c,d} \sum\limits_{^{i,j}_{i \ne j}}^{N}
    V_{cd}(\vec r^{}_{j,d} -  \vec r^{}_{i,c})
  + \frac{m_e}{2} \sum\limits_{i=1}^{N} v_{i,e}^2
  + \frac{m_i}{2} \sum\limits_{i=1}^{N} v_{i,i}^2
\eeq
is conserved.  If the particle trajectories are calculated including
the mean field force, the energy ${\cal E}'_{\rm pot} + {\cal
E}'_{\rm kin}$ is not conserved.  Nevertheless, the conservation law can be
fulfilled by including the mean field energy ${\cal E}_{\rm field} =
L^3 \epsilon_0 E^2/2$ so that the total energy ${\cal E}'_{\rm tot} =
{\cal E}'_{\rm pot} + {\cal E}'_{\rm kin} + {\cal E}_{\rm field}$ is
conserved.  This is illustrated by simulations below.

The occurrence of plasma oscillations can be demonstrated in the
following way. If the mass ratio between electrons and ions $m_i/m_e$ is large
the ion current can be neglected in Eq.~(\ref{current}). After that the
derivative of the total current density is obtained from
\bea\label{jderiv}
& & {d\vec J(t) \over dt}
    = - \frac{e}{L^3} \sum\limits_{i=1}^{N} {d\vec v_i \over dt}
    = \frac{e N}{m L^3} (e\vec E - \vec \xi), \\
& & \vec \xi = \frac{1}{N} \sum\limits_{i=1}^{N} \vec
F_{i,e}^{\,\,\rm short}
    = \frac{1}{N} \sum\limits_{i=1}^{N}
      \sum\limits_{j=1}^{N} \vec F_{ij}.
   \label{xi}
\eea
The force $\vec \xi$ includes only electron-ion interaction forces as
all electron-electron interaction forces are compensated since they
do
not change the total momentum of the electrons.  Although the force
$\vec F_{i,e}^{\,\,\rm short}$ on each electron is typically much
greater than the force $e\vec E$ from the mean electric field, the
average over all electrons is of the same order of magnitude as
$e\vec E$.  If we now differentiate Eq.~(\ref{E-j}) and substitute the
derivative of the current using~(\ref{jderiv}), we obtain the equation for the mean field
\beq
  {d^2\vec E \over dt^2} + \omega_{\rm pl}^2\vec E
  = \frac{\omega_{\rm pl}^2}e \vec \xi.
\eeq
On an average, $ \vec \xi$ vanishes, so that  plasma oscillations
are described. The corresponding oscillations in the current ACF are
obtained from MD simulations as the results below show.

\begin{figure}[ht]
\begin{center}
  \includegraphics[width=0.5\linewidth]{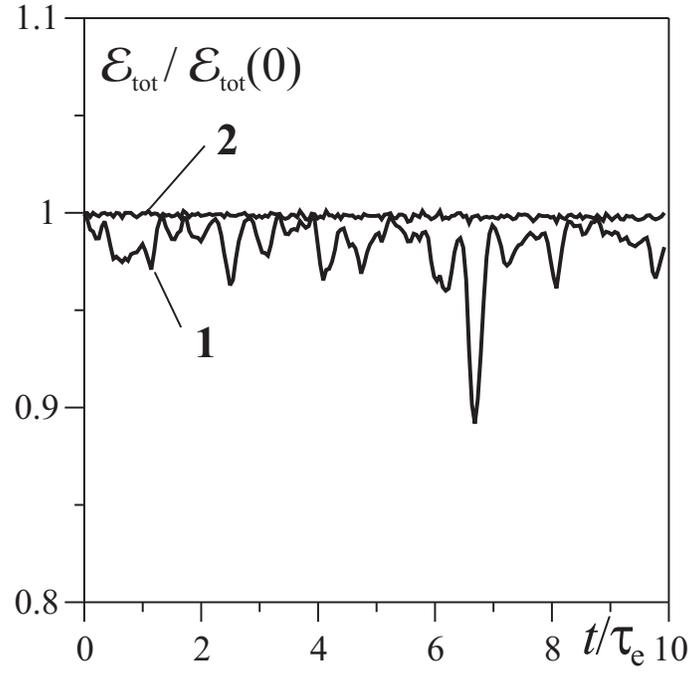}
  \vspace*{-15pt}
\end{center}
\caption{Conservation of the total energy in MD simulations;
curve {\bf 1} --  total energy of the particles
${\cal E}'_{\rm pot} + {\cal E}'_{\rm kin}$ according to
Eq.~(\ref{TotEnergy}), curve {\bf 2} -- total energy ${\cal
E}'_{\rm tot}$ including the mean field energy.
   \label{figen}}
\end{figure}

We now present MD simulations based on the
solution of the equations of motion~(\ref{EqMotion}) in comparison to
the  MD simulations as presented in the previous Sec.~IV where the
contribution of the mean field $ - e\vec E$ was not taken into
account.    The energy conservation is demonstrated in Fig.~\ref{figen} according to
Eq.~(\ref{TotEnergy}).  It can also be seen that the field energy
${\cal E}_{\rm field}$ is rather small compared to the particle
energy
${\cal E}'_{\rm pot} + {\cal E}'_{\rm kin}$.

Results for the longitudinal and transverse current ACF are shown in Fig.~\ref{figjj}. After including
the mean field into the MD simulations, the plasma oscillations in
$K(t)$ become well pronounced in contrast to a monotonously decreasing
behaviour. It should be stressed that the amplitude of these oscillations
does not depend on $N$.

\begin{figure}[ht]
\begin{center}
  \includegraphics[width=0.5\linewidth]{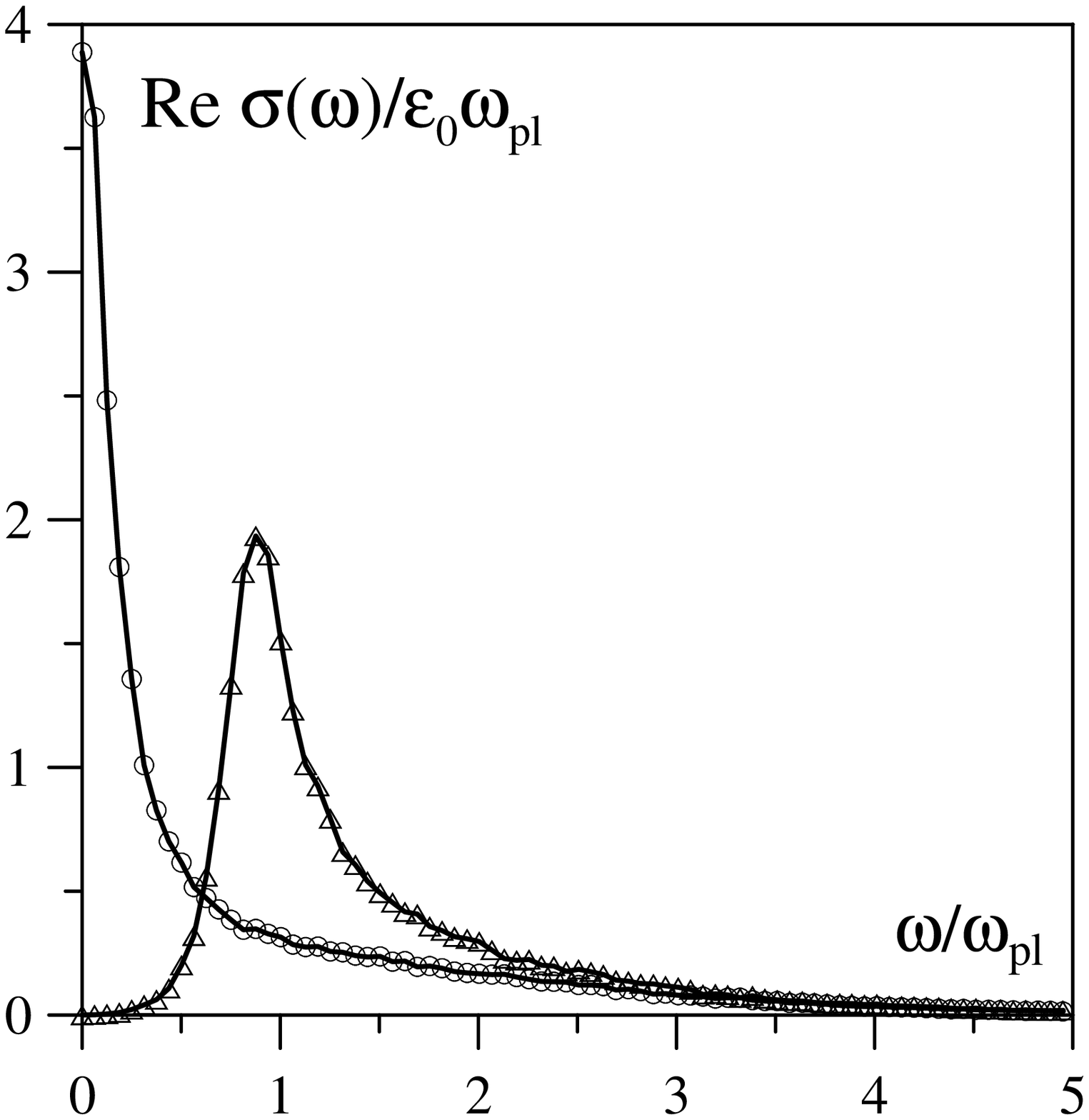}
  \includegraphics[width=0.5\linewidth]{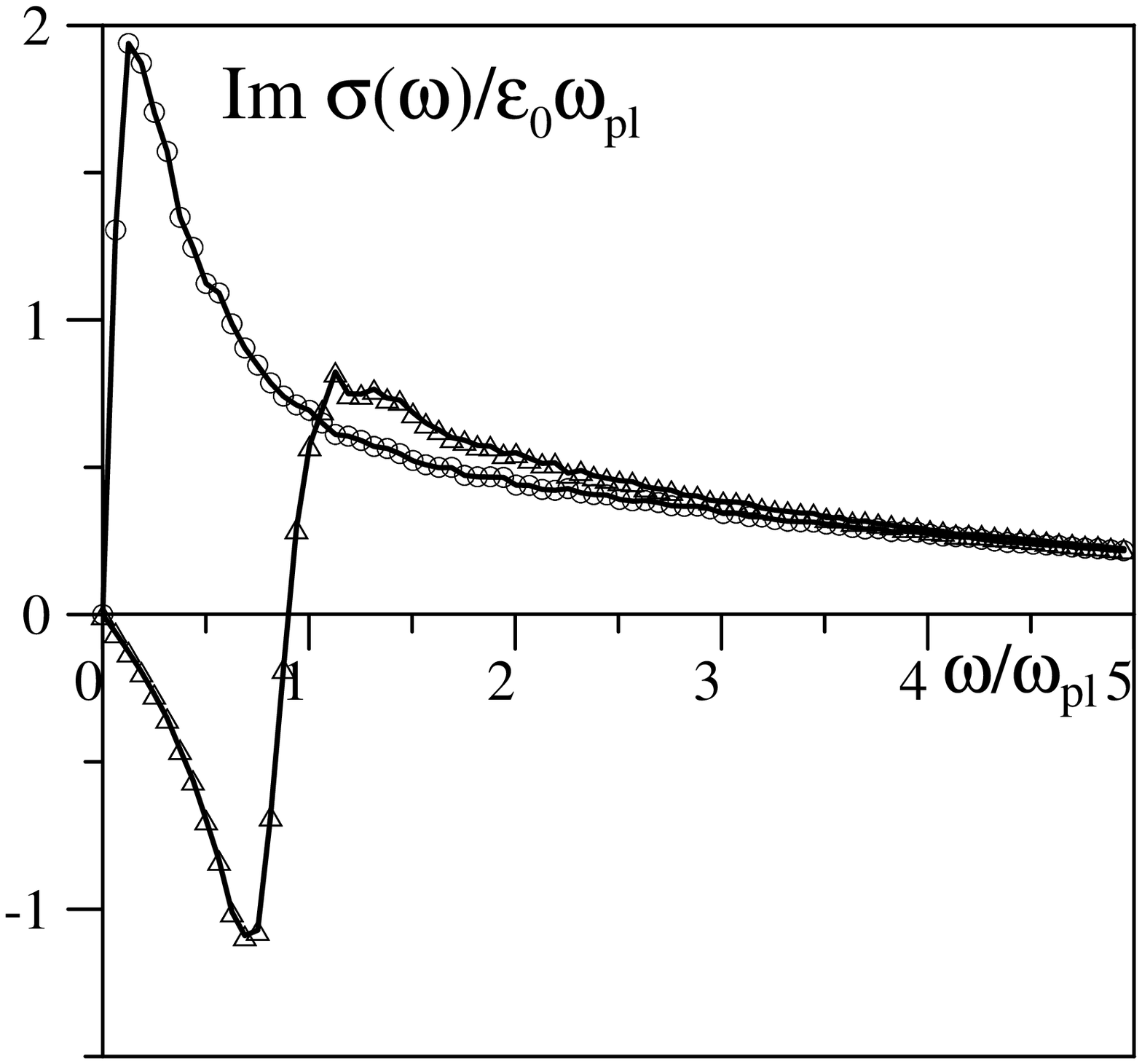}
  \vspace*{-15pt}
\end{center}
\caption{Real and imaginary parts of the Laplace transformation of the
current ACF;  MD simulations without (circles) and including
(triangles) an additional mean-field term in the equations of motion.
  \label{figKw}}
\end{figure}

\begin{figure}[ht]
\begin{center}
  \includegraphics[width=0.5\linewidth]{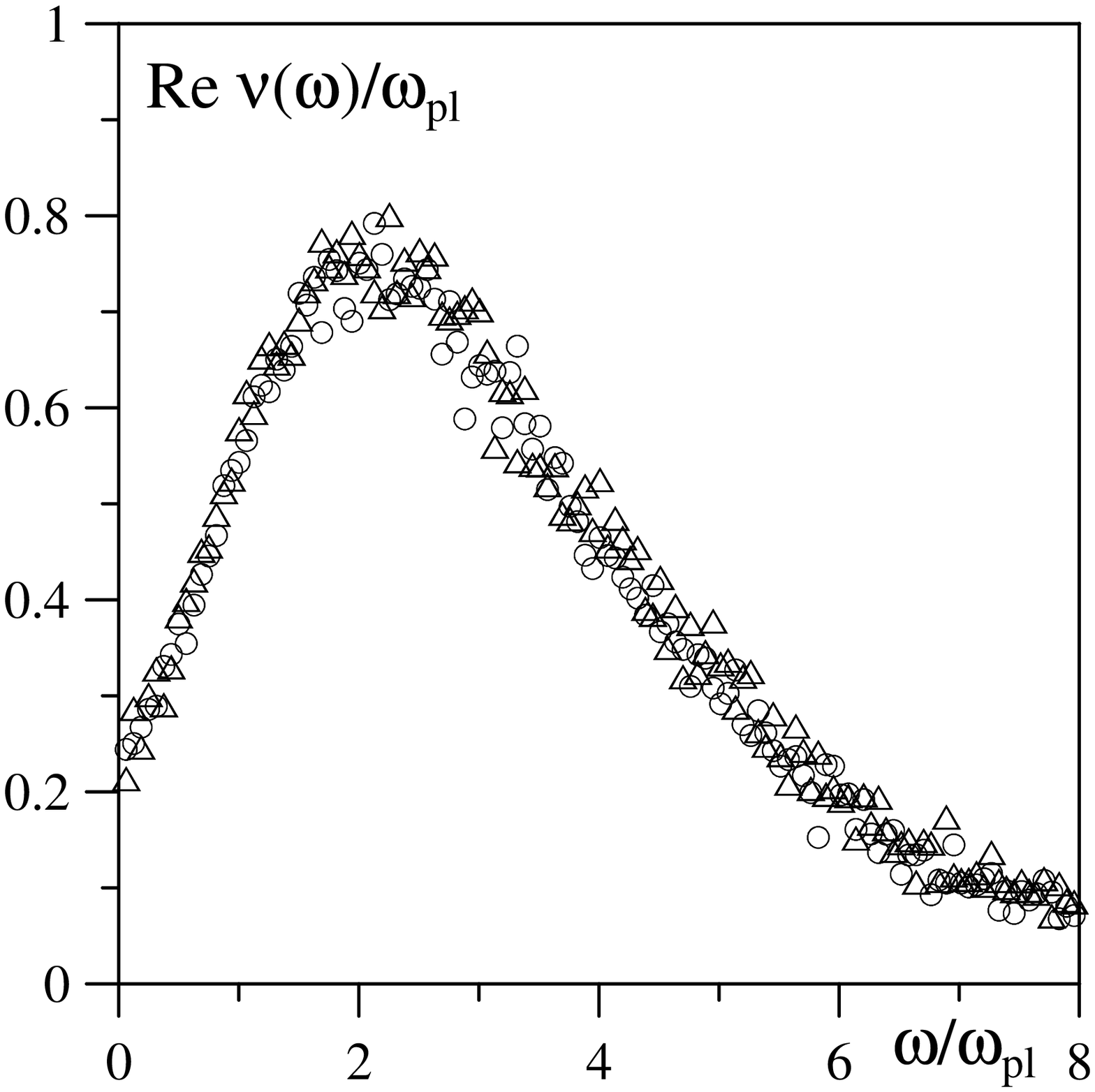}
  \includegraphics[width=0.5\linewidth]{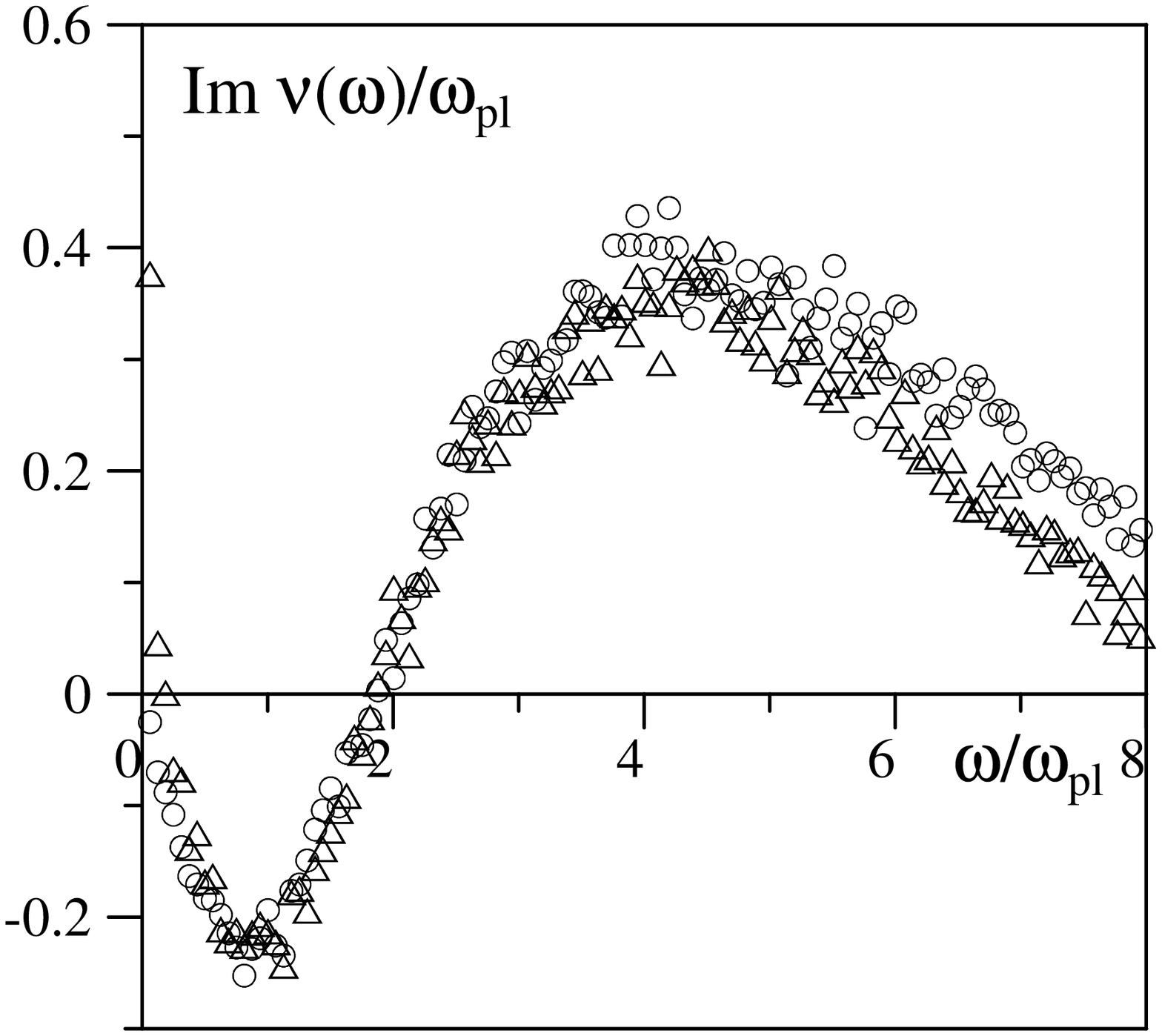}
  \vspace*{-15pt}
\end{center}
\caption{Real and imaginary parts of the dynamic collision
frequency or memory function from   MD simulations without (circles)
and including (triangles) an additional mean-field term in the
equations of motion.
  \label{figNu}}
\end{figure}

The conductivity calculated  according to Eq.~(\ref{sigmaMD}) is shown in
Fig.~\ref{figKw}.  In comparison to the transverse case, the conductivity shows a qualitatively different behaviour.  The real part following from the MD simulations including mean field is zero for zero frequency as is expected from the
expression for the external conductivity (\ref{KDrude}). For the case
without  mean field, ${\rm Re} \,\sigma$ has a finite value.  In the
high frequency limit, both curves coincide.  The dynamical collision
frequencies $\nu(\omega)$ and the memory function $\widetilde\nu(\omega)$ calculated from the simulation data for the ACFs are shown in Fig.~\ref{figNu}.  As
pointed out, the results for the Laplace transform of the ACF differ significantly
(Figs.~\ref{figKw}).  Nevertheless, if  Eq.~(\ref{nukap}) is used for the collision frequency $\nu(\omega)$ and Eq.~(\ref{wrongDrude}) for the memory function
$\widetilde\nu(\omega)$ in order to calculate the collision frequency,
the results for both coincide quite clearly (Fig.~\ref{figNu}).

Therefore, our analysis showed that
the contradiction between the transverse conductivity which should be identical with the internal conductivity in the long-wavelength limit and the external conductivity
could be resolved if the mean field is taken into account.  The
difference between ${\rm Im}\,\nu(\omega)$ and ${\rm
Im}\,\widetilde\nu(\omega)$ in the low frequency limit is caused by
the numerical error of ${\rm Im}\,\nu(\omega)$ due to substraction
of two large terms in Eq.~(\ref{nukap}).

\section{Conclusion}

Molecular dynamics simulations of strongly coupled plasmas were
performed using the quasiclassical Kelbg interaction potential.  The
current auto-correlation function was computed for a non-degenerate two-component
plasma. Whereas for finite $k$ the dynamical structure factor and the plasma oscillations are reproduced by MD simulations, see
\cite{Hansen81a,Hansen87,Zwicknagel}, the original methods do not allow to
consider $k$ values with $k<2\pi/L$. On the other hand $k=0$ should
be possible to investigate with MD simulations in a finite volume.

We presented calculations for the transverse current ACF as well as for the longitudinal one. Although in the limit $k \rightarrow 0$ the transverse and longitudinal dielectric function and conductivities, respectively, coincide, the current ACF behave differently in this limiting case.
It was shown that the results of MD simulations without a mean field in the
long wavelength limit provide the monotonously decreasing transverse ACF.
Its Laplace transform is to be directly related to the transversal
conductivity. 

In MD simulations for the longitudinal case, a mean-field term has to be included
into the equations of motion in addition to the short range forces
inside the Debye sphere.  This mean-field term originates from
surface charges not taken into account in the usual procedure of
force calculation by the nearest image method.
Simulations with these altered equations
of motion show well pronounced plasma oscillations in the longitudinal current ACF. The results for the collision frequency as
obtained in both simulation methods using the corresponding relations
for the internal or external conductivities do coincide.  

Additionally, the dynamical collision frequency inferred from the simulation data was compared with analytical
results, which were derived using  a generalized linear response theory.
We found  good agreement in the low and high frequency limits for a
moderate nonideality. In particular, for $\omega<\omega_{\rm pl}$,
classical MD simulations using the corrected Kelbg potential are able
to reproduce the quantum behaviour of Coulomb plasmas.

\section{Acknowledgements}
The authors are thankful to G.E.~Norman, A.A.~Valuev and G.~Zwicknagel for
fruitful discussions. I.M. acknowledges the support from
RFBS by grant 03-07-90272v, Integracia by grants U0022, I0661, the Dynasty
Foundation and the International Center of Fundamental Physics in Moscow.
H.R. received a fellowship from the DFG and T.M. was supported by the
SFB 198.


\vspace*{30pt}

\end{document}